\documentclass[aps,apl,twocolumn,showpacs,superscriptaddress]{revtex4}
\usepackage{amsmath}
\usepackage{dcolumn}
\usepackage{epsfig}
\usepackage{graphicx}
\usepackage{latexsym}

\def\S2RO4{Sr$_2$RuO$_{4}$}
\def\SRO3{SrRuO$_{3}$}
\def\STO3{SrTiO$_{3}$}

\begin{document}
\hyphenation{Ka-pi-tul-nik}

%\twocolumn[
%\hsize\textwidth\columnwidth\hsize\csname@twocolumnfalse\endcsname

%\draft

\title{Modified Sagnac interferometer for high-sensitivity magneto-optic measurements at cryogenic temperatures}

\author{Jing Xia}
\affiliation{Department of Physics, Stanford University, Stanford, CA 94305}
%\email{xiajing@stanford.edu}
\author{Peter T. Beyersdorf}
\affiliation{Department of Physics and Astronomy, San Jose State University, San Jose, CA 95192}
\author{ Martin M. Fejer } 
\affiliation{Department of Applied Physics, Stanford University, Stanford, CA 94305}
\author{Aharon Kapitulnik}
\affiliation{Department of Applied Physics, Stanford University, Stanford, CA 94305} 
\affiliation{Department of Physics, Stanford University, Stanford, CA 94305}

\date{\today}

\begin{abstract}
We describe a geometry for a Sagnac interferometer with a zero-area Sagnac loop for measuring magneto-optic Kerr effect (MOKE) at cryogenic temperatures. The apparatus is capable of measuring absolute polar Kerr rotation at 1550 nm  wavelength without any modulation of the magnetic state of the sample, and is intrinsically immune to reciprocal effects such as linear birefringence and thermal fluctuation. A single strand of polarization-maintaining (PM) fiber is fed into a liquid helium probe, eliminating the need for optical viewports. This configuration makes it possible to conduct MOKE measurements at much lower temperatures than before. With an optical power of only 10 $\mu$W, we demonstrate static Kerr measurements with a shot-noise limited sensitivity of  $1\times 10^{-7}$ rad/$\sqrt{\rm Hz}$ from room temperature down to 2K. Typical bias drift was measured to be $3\times 10^{-7}$ rad/hour.
\end{abstract}

%\pacs{74.72.Hs, 74.50.+r, 74.25.-q}

\maketitle

Magneto-optical effects result from the interaction of photons with spins, mediated through spin-orbit coupling, and provide a broad variety of tools for investigation of magnetic and electronic properties of materials.  Linearly polarized light that interacts with magnetized media can exhibit both ellipticity and a rotation of the polarization state \cite{qiu}. These effects are generally catagorized into two primary phenomena, the Faraday Effect which occurs when electro-magnetic radiation is transmitted through a magnetized medium, and the Kerr Effect which describes the state of the light reflected from the magnetized medium. This second phenomenon also called the  Magneto-Optical Kerr Effect (MOKE)  \cite{petros}, is further catagorized by the direction of the magnetization vector with respect to the reflection surface and the plane of incidence. In magnetic materials the size of MOKE signals is large enough to be detected with simple crossed-polarizer techniques. To achieve higher sensitivity, modulation schemes are needed. For example,  it has recently been demonstrated \cite{kato} that better than 1 $\mu$rad sensitivity can be achieved with lock-in detection when probing oscillating magnetizations. However,  there are many cases in which static magnetization needs to be studied, or no modulation scheme can be employed (i.e. a DC mode). Of particular interest is the search for  time-reversal-symmetry-breaking (TRSB) states \cite{mackenzie} in unconventional superconductors such as  \S2RO4 ($T_c \approx 1.5$ K). In searching for TRSB effects measurements must be performed at low temperatures ($T \ll T_c$) which requires very low power while at the same time there is no way to modulate the orientation of the chirality in the material. Finally, the ability to measure Kerr rotation with high sensitivity at small optical power is especially important in near field MOKE microscopy where the reflected intensity is very weak and modulation is not always possible.

Many of the difficulties associated with DC MOKE measurements can be largely overcome by the use of fiber Sagnac interferometry \cite{spielman1,spielman2,kdf}, which measures the relative phase shift between two beams of light that travel an identical path in opposite directions.    2 $\mu$rad/$\sqrt{\rm Hz}$ Kerr rotation sensitivity and  1 $\mu$rad/$\sqrt{\rm Hz}$ Faraday rotation sensitivity have been demonstrated \cite{kdf} with this technique. The Sagnac interferometer is capable of sensing very small nonreciprocal effects, such as magneto-optic effects and mechanical rotation of the loop, while rejecting reciprocal effects such as linear birefringence, and is thus well suited for probing static magnetizations. Near-field Sagnac magneto-optic microscopy has also been proposed \cite{kdf} and realized \cite{petersen,meyer}. However, the rather complicated design limits its uses in magneto-optics; like other MOKE techniques, it is not really suited for low temperature applications due to the need of optical viewports.

Here, we report a modified fiber-Sagnac interferometer for MOKE measurements. The greatly simplified design not only enables a DC sensitivity of $1\times 10^{-7}$ rad/$\sqrt{\rm \rm Hz}$ with only 10 $\mu$W optical power, but also is well suited for low temperature measurements by eliminating optical viewports. We used this setup for Faraday measurements as well by placing a non-magnetic mirror behind the transparent sample. To minimize DC bias drifts, we chose an in-loop modulation \cite{spielman1,spielman2,kdf} scheme instead of post-modulation \cite{beyersdorf1}; the reason can be seen later. 

\begin{figure}[h]
\begin{center}
\includegraphics[width=1.0 \columnwidth]{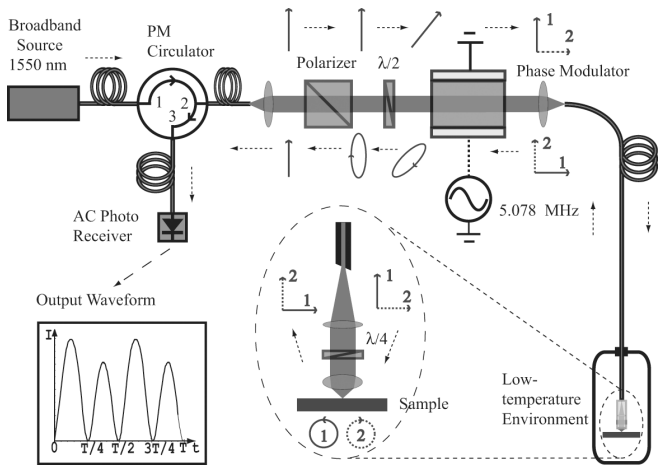}
\end{center}
\vspace{-4mm}
\caption{ Experimental setup and the polarization states at each locations: vertical, in-plane polarization; horizontal, out-of-plane polarization; solid line, beam 1; dashed line, beam 2. } 
\label{setup}
\end{figure}

Figure \ref{setup} shows a schematic of the design. The beam of light from a broad-band light source centered at 1550 nm is routed by a fiber polarization-maintaining (PM) circulator to a Glan-Taylor polarizer, which polarizes the beam. The PM circulator transmits light from port 1 to 2 and port 2 to 3 with better than 30 dB isolation in the reverse direction. A half-wave ($\lambda /2$) plate is placed after the cube polarizer to conveniently rotate the polarization of the beam at 45$\rm ^o$  to the axis of a electro-optic modulator (EOM), which generates a $5.078$ M\rm Hz time-varying phase shift $\phi(t) = \phi_m \sin (\omega_m t)$ on the in-plane polarization. The $\rm LiNbO_3$ crystal used in the EOM is highly birefringent and the input broadband linearly polarized beam is split into two incoherent parts with equal power after passing through the EOM: beam 1 with in-plane polarization is modulated and beam 2 with out-of-plane polarization is un-modulated. The two beams are then launched into the fast and slow axes respectively of a  10-m-long PM fiber that is fed into a cryostat. Upon exiting the fiber, the two orthogonally polarized beams are converted into right- and left-circularly polarized light by a quarter-wave ($\lambda /4$) plate, and are then focused by a NA = 0.3 aspheric lens onto the sample. The non-reciprocal phase shift $\phi_{nr}$ between the two circularly polarized beams upon reflection from the magnetic sample is twice the Kerr rotation \cite{spielman1,spielman2,kdf} ($\phi_{nr} = 2 \theta_K$). The same quarter-wave plate converts the reflected beams back into linear polarization, but with a net 90$^o$ rotation of the polarization axis. Namely, beam 1, which was in-plane polarization, is now out-of-plane and is launched into the slow axis of the PM fiber; while beam 2, which which was input along the slow axis of the fiber, now goes back along the fast axis. The two beams then pass through the EOM again but beam 2 is modulated this time. At this point, the two beams have gone through exactly the same path but in opposite directions, except for the phase difference of  $\phi_{nr} = 2 \theta_K$ from reflection off of the sample and a difference $(\tau)$ in the time when they were modulated by the EOM. The two beams are once again coherent, and interfere and produce an elliptically polarized beam, whose in-plane component is routed by the circulator to a  125-M\rm Hz-bandwidth photoreceiver. Lock-in detection was used to measure both first ($I_\omega$) and second ($I_{2\omega}$) harmonics of the modulated intensity $I(t) = |E_{in-plane}(t)|^2$. The modulation frequency ($\omega_m$) is tuned so that the time delay is $\tau = \pi/\omega_m$ . It can be shown \cite{spielman1} that in this case the harmonics have simple forms:  $I_\omega = I_0( \sin (2\theta_K) J_1(2\phi_m))$ and $I_{2\omega} = I_0(\cos (2\theta_K) J_2(2\phi_m))$, where $I_0$ is the total intensity.  The Kerr rotation can then be calculated as

\begin{equation}
\theta_K = \frac{1}{2} \tan^{-1}\left[ \frac{J_2(2\phi_m)I_\omega}{J_1(2\phi_m) I_{2\omega}}\right]
\end{equation}

\noindent where $J_1$ and $J_2$ are Bessel functions.

It can be easily seen that the area of the Sagnac loop is zero, as the incident and reflected beams travel along the same PM fiber but with orthogonal polarizations. This not only eliminates bias drifts due to mechanical rotation \cite{mackenzie} of the loop, but also largely suppresses the nonlinear optical Kerr effect which was the dominant source \cite{dodge} of offsets in previous fiber Sagnac interferometers, especially in the reflection configuration \cite{dodge}. DC bias drifts due to misalignments of the polarization axes of optical components are also largely suppressed because of the following considerations: First, although misalignment between the bulk polarizer and the EOM could mix the in-plane and out-of-plane components of the elliptically-polarized beam returning to the polarizer, the ratio between the first and second harmonics won't change and thus this misalignment will at the best serve as a loss in the total detected intensity. Second, misalignment between the fiber axes and quarter-wave ($\lambda /4$) plate, as well as imperfections in the quarter-wave plate itself and the finite extinction of the PM fiber (especially at low temperatures) could couple light into the wrong polarization. However, the high birefringence of the EOM crystal and of the PM fiber along with the short coherence length ($\approx 30 \mu$m) of the broadband source renders these competing beams incoherent with the two major beams and minimizes the resulting offsets. In practice, the bias offset in our system drifts $\pm 1.5 \mu$rad over an interval of 24 hours.

\begin{figure}[h]
\begin{center}
\includegraphics[width=0.8 \columnwidth]{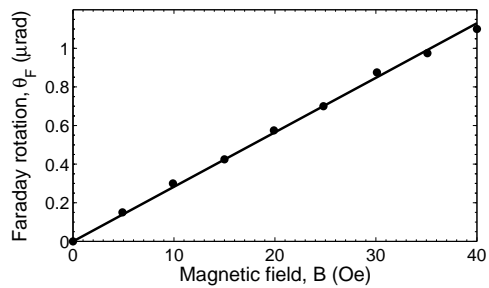}
\end{center}
\vspace{-4mm}
\caption{ Faraday rotation of 0.5 mm thick fused silica slab as a function of the applied magnetic field. Measurements were done at room temperature. } 
\label{silica}
\end{figure}

This apparatus has been calibrated for both large and small magneto-optic effects. We measure the saturated Faraday rotation of a $40 \pm 2 \mu$m thick bismuth-doped rare-earth iron garnet crystal to be $3.3^o$ ($83 \pm 5$ deg/mm), which is in reasonable agreement with the  90 deg/mm nominal value. We also calibrated the apparatus in the sub-$\mu$rad  range by measuring the Faraday rotation of a $500-\mu$m-thick slab of fused silica. The result is shown in figure \ref{silica} and the measured Verdet constant was $5.52 \times 10^{-8}$ rad/Oe-mm, which agreed to within 5$\%$ with literature values \cite{rose,cruz}. To ensure that our apparatus is immune to linear birefringence, we also tested on some highly birefringent samples: quarter-wave-plates designed for 840 nm and 670 nm wavelengths. Both tests showed null results.

\begin{figure}[h]
\begin{center}
\includegraphics[width=1 \columnwidth]{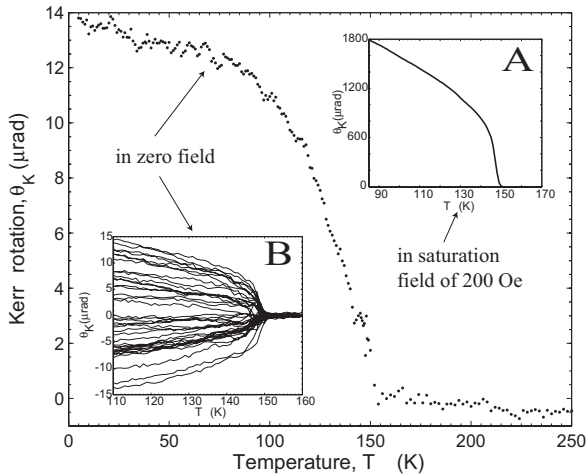}
\end{center}
\vspace{-4mm}
\caption{Polar Kerr effect from a \SRO3  thin film. The 30 nm film was cooled in zero field through its 148 K Curie temperature. Insert A: Kerr rotation of the same sample cooled down in a saturation field of 200 Oe; Insert B: Kerr rotations of the same sample measured in different cool-downs in zero fields. }
\label{srruo3}
\end{figure}

As a test of the apparatus at cryogenic temperatures, we examined the ferromagnetic phase transition of a  30-nm-thick expitaxial film of  \SRO3 on  \STO3 substrate. The film was oriented with the  [100] direction perpendicular to the plane of the film and has a transition temperature $T_c \approx 148 \rm K$. Figure \ref{srruo3}  shows the measured Kerr rotation as a function of temperature in zero magnetic fields. When cooled in zero fields, the magnetizations orient randomly from domain to domain. The optical beam was focused on the sample with a $N.A. \approx 0.3$ objective and had a diameter of $w \approx 3 \mu$m , while the sizes of the magnetic domains are much smaller (on the order of film thickness). Thus the measured Kerr rotation is the averaged effect of domains within the optical spot, and should fluctuate around zero. To verify this, we measured the Kerr rotations upon warming and cooling in zero fields and the results are shown in insert B of figure \ref{srruo3}. Kerr rotations during different cooldowns fluctuate from positive to negative as expected and have a standard deviation of $\sigma = 8.5 \mu$rad at 110 K. To estimate the domain size, we measured the pure Kerr rotation by cooling the sample in a saturation field of 200 Oe. We found the saturated Kerr rotation at 110 K to be  $\phi_0 = 1441 \mu$rad  (figure \ref{srruo3}, insert A).  Assuming random domains, we can then estimate the domain size to be $d = w \sigma/\phi_0 \approx 18 nm$, which is of the same order of the film thickness as expected. However, further investigation of magnetic domains requires methods with higher spatial resolution and near field scanning can be a good choice.

In conclusion, we have demonstrated an optical-viewport-free magneto-optic Sagnac interferometer that is well suited for low temperature MOKE measurements with high sensitivity and low optical power. The much simplified design compared to previous magneto-optic Sagnac interferometers could broaden its uses in MOKE studies.

\acknowledgments
We would like to thank Wolter Siemons and Gertjan Koster for depositing the \SRO3  sample, as well as Integrated Photonics Inc. (Hillsborough, NJ) for providing the bismuth-doped rare-earth iron garnet crystal which was used for calibrating the apparatus. This work was supported by Center for Probing the Nanoscale, NSF NSEC Grant 0425897.

\end{document}